\documentclass[aps,prb,preprint,graphicx,epsf,amsbsy,amssymb]{revtex4}
\everymath{\rm}
\everydisplay{\rm}

\input {amssym.def}
\input {amssym.tex}

\usepackage{graphicx,epsf}

\begin{document}
\title{Angle Resolved Photemission Experiments to look for Time-reversal Violation in
Cuprates}
\author{	J. C. Campuzano$^{1,2},$
            A. Kaminski$^{1,3}$ and
            C. M. Varma $^{4}$} 
\address{$^1$University of Illinois at Chicago\\ Chicago IL 60607, USA \\
			$^2$Argonne National Laboratory\\ Argonne IL  60439, USA\\
            	$^3$University of Wales Swansea\\ Swansea SA2 8PP, UK\\
                  $^4$Bell Laboratories, Lucent Technologies\\ Murray Hill, NJ 07974, USA\\}
\begin{abstract}
 The recent experiments reported by Borisenko et al. \cite{BORISENKO}, are examined in light of the conditions to be satisfied in the search for time-reversal violation by circularly polarized ARPES. Two principal problems are found:
(1) A lack of any evidence for the magnitude of the pseudogap or the temperature
of its onset in the samples studied. 
(2) A difference in the dichroic signal at low and high temperatures. The difference is greater than the stated error bars and is contrary to the conclusions reached in the paper. 
\end{abstract}

\maketitle

A new experimental technique using circularly polarized Angle-resolved photoemission (CPARPES)  
was suggested \cite{SIMONVARMA} as a way to search for the proposed Time-Reversal Violation
in the pseudogap phase of the high temperature superconductors.
The essentials of the technique are the following: 
The magnetic-field averaged over 
a unit-cell is zero 
as it changes direction in 
specified patterns in a unit-cell. 
However, any given single-particle excitation has a wavefunction of the form:
\begin{equation}
\Psi_{{\bf k}}(\{r\})=  \Psi_{e,{\bf k}}(\{r\})+i\theta  \Psi_{o,{\bf k}}(\{r\}).
\end{equation}
where both $\Psi_e$ and $\Psi_o$ are real. Time-reversal is violated because
 $\Psi_{{\bf k}}(\{r\})$ cannot be transformed to a real wave-function by any 
gauge transformation. Further $\Psi_{e,{\bf k}}(\{r\})$ is even about
 some-specified mirror planes while $\Psi_{o,{\bf k}}(\{r\})$ is odd about 
the same mirror planes of the {\it lattice}. So $\Psi_{{\bf k}}(\{r\})$ 
 do not have reflection symmetry about certain mirror planes 
of the {\it lattice} \cite{SIMONVARMA}. CPARPES experiments couple 
the external, circularly polarized photons to the current in the lattice and are 
linearly sensitive to the wave-functions. There will therefore be a difference in
intensity of the emitted electron current for left (lcp) and right (rcp) circularly polarized 
photons if there is a violation of reflection symmetry about the specified mirror planes. 
This difference in intensity will, in general, be proportional to $\theta$. However an 
experiment sensitive to the charge density, such as an ordinary diffraction 
experiment or linearly polarized ARPES, will show no reflection symmetry breaking -
 at least to order $\theta$.

 The essence of the experiment is to observe the 
violation of reflection symmetry about certain mirror planes as one cools
 through the characteristic pseudogap temperature in a CPARPES experiment, 
while at the same time observing no similar violation in experiments such
 as ordinary linearly polarized ARPES \cite{KAMINSKI}.
 
There is however one important technicality, which actually helps 
in the design of the experiment and was employed in 
the experiment \cite{KAMINSKI} that observed the predicted effect. Let $\hat{m}$ be a 
vector in a mirror plane,  $\hat{n}$, the direction of polarization of 
the incoming photons and ${\bf k}$  be the momentum of the electronic
 wave-function in the lattice (which in the first Brillouin zone has the 
same direction as that of the outgoing electron). If these three 
vectors are non co-planar, i.e. if $\hat{m}\cdot(\hat{n}\times{\bf k})  \neq 0$, 
the intensity of the current of the outgoing electrons for right circularly 
polarized (rcp) and left circularly polarized (lcp) photons are unequal 
even when Time-reversal/Reflection symmetries are unbroken. 
Suppose $\hat{n}$  is in the mirror plane, then the difference of the rcp and lcp
 intensity, D, is zero if ${\bf k}$ is collinear with $\hat{m}$. Further, D is odd with 
respect to the component of  $\bf{k}$ perpendicular to $\hat{m}$. We call 
this the {\it geometric} effect. 
Assume now that Time-reversal is violated below the pseudogap temperature with reflection symmetry violated about a specified $\hat{m}$. There is now an extra component to D (proportional to $\theta$) which is finite even when 
 ${\bf k}$ is collinear with $\hat{m}$. Furthermore D is even with respect to the component of  $\bf{k}$ perpendicular to $\hat{m}$. Taken together, the geometric effect and the  time-reversal violation effect lead to a finite value of D at 
the mirror plane  below the pseudogap temperature, while it remains zero above. 
The essence of the experiment is to measure D as a function of temperature
 and as a function of  ${\bf k}$. The  point on the momentum axis for which D is zero should appear to move when the sample is cooled below the pseudogap temperature. 

Besides the requirement of high momentum resolution, it is essential to measure absolute intensities 
in order that D can be determined with sufficient accuracy. Otherwise only an upper limit can be put on D.
It is of course also necessary to know the pseudogap temperature $T^*$ and magnitude for the samples studied. These quantities can be measured via ordinary ARPES experiments because the proposed dichroism effect only occurs below $T^*$ and its magnitude is related to the magnitude of the pseudogap.
  
In one recent experiment \cite{BORISENKO} to search for the 
signature of Time-reversal violation using CPARPES it was claimed that no time reversal violation occurs.
However, this work did not fulfill the above conditions. The experiments were done in Pb-doped BISCCO, 
(contrary to the much studied BISCCO compounds). 
The first and most important 
problem with the Borisenko et al. paper is that there is no
 evidence presented - or in any papers referenced - regarding  
  the  magnitude of the pseudogap or $T^*$, as measured by photoemsission 
or transport experiments. A comparison
 of Fig.~1e, which shows the energy distribution curve of an  overdoped sample 
 with Figs.~3a and 3b for a supposedly underdoped sample, 
does not show any changes characteristic of
the pseudogap in the latter. Such differences are now routinely observed by other experiments in underdoped BISCCO. 
At the very least the pseudogap in these Pb-doped samples is significantly smaller than those 
in underdoped BISCCO to render it unnoticeable when comparing 
Figs. 1 and 3.

\begin{figure}
\includegraphics[width=3.5in]{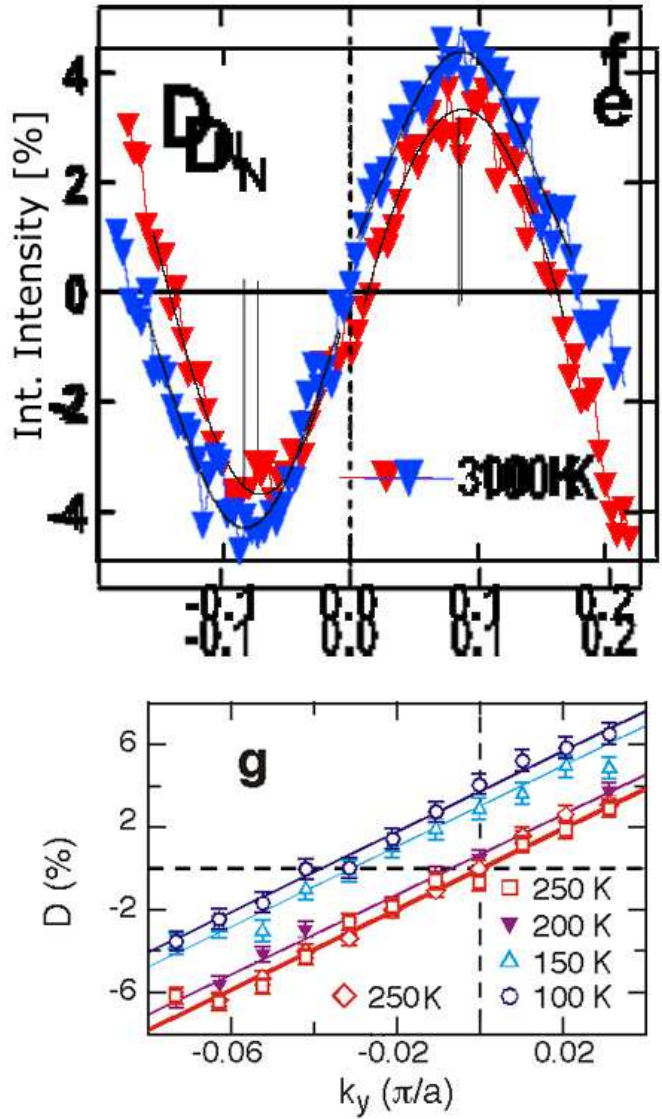}
\caption{\label{fig1} Comparison of the data for an underdoped sample
 from the two CPARPES experiments. 
(a) Superimposed dichroism signal data for T=100K (red triangles)
 and T=300K (blue triangles)  from fig. 3e and 3f in Ref.~[1]. 
Please note: the axis were carefully aligned to match the scales in the figures.
(b) Dichroism data at T=100K, 150K, 200K, 250K for underdoped sample from 
 fig. 3g in Ref.~[3]. Please note different momentum range here.
}
\end{figure}

Keeping this in mind, we now look at the CPARPES data presented 
in Ref.~[1] and re-examine its conclusions. A proper analysis requires one to know the error bars 
in the experiment. At one point in the paper, it is stated that
 the mirror planes are determined to an accuracy of 0.001$A^{-1}$.
 At another point it is stated that the accuracy in determining the momentum 
is $\pm 0.004 A^{-1}$. 
We have taken the data from Ref.~[1] in Fig. 3e
 at 300K and 3f at 100K and plotted them both in fig. 1a here. 
It is quite clear that near the mirror plane there is a 
systematic shift of the two curves. The average difference of the mid-points
of the red and the blue triangles between $\pm 0.05 A^{-1}$
is found to be $0.012 A^{-1}$. 
Either the error estimates of the authors are incorrect or the anticipated effect has been observed 
to an accuracy of about 3 standard deviations. 
The magnitude of the effect in $A^{-1}$, if the given error bars 
given are to be taken seriously, is about $1/2$ that observed earlier in Ref.~[3] 
(shown here in Fig.~1b). This is not surprising for three reasons:
1) if a pseudogap exists in the Pb-doped sample, it is much smaller 
than those observed in underdoped samples of BISCCO, as discussed above. 
2) The overall magnitude of the difference 
signal due to the geometric effect at room temperature, i.e. above T* is also about 1/2 that observed earlier \cite{KAMINSKI}.
This smaller effect could be due to a difference in the degree of circular polarization about 
which no information is given in Ref.~[1]. 
3) As stated in Ref.~[3], the results presented there represent the maximum dichroism signal seen in 
any recycling experiment, the variation due presumably to changing domains.

One may also look at two other relevant sets of data in Ref.~[1]. 
Fig.~3c shows a systematic average shift of $0.008 A^{-1}$, twice the quoted error bar,  
in the two sets of data presented. 
Fig.~3d gives the data only at 100K and 30K. 
There is indeed no shift of the two, just as there is none discernible in Fig.~3g of 
Kaminski et al.\cite{KAMINSKI} below 150K. In the absence of higher temperature data,
 Fig.~3d does not say very much. One may note however that the
 intensity of the two maxima in the curves differ by about 2\% ,
 whereas in the absence of an effect, they should be equal within the 
stated error bar of 0.05\%. 
 The three figures discussed above represent all the data relevant to dichroism
near the $(\pi,0)$ and equivalent points; all of them show evidence of an effect that is 
well outside the quoted error bars.

The casual attitude towards data analysis is further in evidence 
in Fig.~4b, where the 0.03$A^{-1}$ position of the point of zero 
$D_N$ in Fig.~4b is considered equivalent to 0 within error bars.

We take this opportunity to also comment on the
"independent experimental approach" which has been used. 
To quote the paper "...the signal is weak and problematic 
for the application of the absolute intensity criterion", 
the criterion used by Kaminski et al. \cite{KAMINSKI}. Therefore Borisenko et al.
employ an alternative
"lineshape analysis criterion". This appears to be
based on the misconception that when Time-reversal violation occurs, the mirror plane rotates. 
The "lineshape criterion' is employed as follows: 
D is measured along the zone diagonals. Within the error bars
 it changes sign along this line as expected and its absolute 
magnitude saturates and falls off as the two Fermi-surface points 
f1 and f2 are approached. The "lineshape criteria" consists of fits to the data, which appears as
an antisymmetric curve. The mid-point of the maxima and minima in these fits determine the 
mirror planes. 
If within the error bars of the momentum resolution, the mid-point, i.e. the position 
of the mirror plane, does not move with temperature, it is concluded that there 
is no time-reversal violation. 
However, the position of the maxima should {\it not} move due
to time-reversal violation. Only the value of the whole 
curve $D({\bf k})$, including at f1 and f2 should
move with respect to that without time-reversal violation. 
The constancy in the position of the maxima is evidence 
only that the {\it lattice} or charge density symmetry does 
not change noticeably. In other words, the experiment cannot
 be done without the "absolute intensity criterion".

 This misconception can have a serious consequence in the analysis:
 It is calculated from Eq. (1) and (2) of the paper of Borisenko et al. that the differences in intensity of 
 incoming lcp and rcp photon intensity "only rescales and shifts
 $D_N$ along the vertical axis leaving the opportunity to define
 ${\bf k}$ locations by the lineshape analysis precisely."
 Actually the shifts in $D_N$ are precisely what one should be searching for.
Any significant difference of the lcp and the rcp photon intensity
can therefore lead to misleading conclusions.

In conclusion, in the absence of a proper characterization of the sample, for example by direct 
 observation of a pseudogap through ordinary linear 
polarized ARPES, the lack of correspondence between stated errors and 
the observed variations in the data and a misunderstanding of what features 
of the data are sensitive to time-reversal violation, make the conclusions of   
Borisenko et al. of dubious validity.

\end{document}